\documentclass[twoside,slac_one]{revtex4}
\usepackage{graphicx}
\usepackage{fancyhdr}
\usepackage{amsmath} 
\usepackage{bm}
\usepackage{amsxtra}
\usepackage{amssymb}
\usepackage{amsthm}
\usepackage{latexsym}
\usepackage{lscape}

\usepackage{subfigure}

\begin{document}

\title{Soft QCD measurements in the forward direction with the LHCb experiment}

\author{A. Camboni, on behalf of the LHCb collaboration}
\affiliation{Departament de Estructura i Constituents de la Mat\`eria, Universitat de Barcelona, Barcelona, Spain}

\begin{abstract}
LHCb presents studies of particle production in minimum bias events in \textit{pp} collisions at $\sqrt s=7$ TeV. These studies include measurements of strangeness production, particle ratios, baryon-antibaryon ratios and charged particle production. The forward coverage and low $p_T$ acceptance of the experiment makes these measurements very complementary to those performed by the central detectors at the LHC. Further benefits arise from the powerful particle identification capabilities provided by the LHCb RICH system. The measurements are compared with theoretical predictions.
\end{abstract}

\maketitle

\thispagestyle{fancy}

\section{Introduction}

LHCb is one of the four large experiments operating at the Large Hadron Collider (LHC) at CERN. It is devoted to precision measurements of CP violation and rare B meson decays. Since the $b\bar{b}$ production in $pp$ collisions at 7 TeV is strongly favored in the forward/backward region, LHCb is build as a single arm magnetic dipole spectrometer with a polar angular coverage with respect to the beam line of approximately 15 to 300 mrad in the horizontal
bending plane, and 15 to 250 mrad in the vertical non-bending plane. A full description of LHCb may be found in~\cite{lhcb_descr}. All subdetectors were fully operational and in a stable condition for the data that are analysed. For the measurements presented here the tracking detectors and particle identification (PID) strategy are of particular importance.

The tracking system is composed by a precision Vertex Locator (VELO) surrounding the interaction region, a dipole magnet and three downstream tracking stations. The resolution for primary (secondary) vertices is $\sim50(100)\ \mu$m and the momentum resolution for tracks having hits in both the VELO and the tracking stations is around 5\%. Two Ring Imaging Cherenkov (RICH) detectors provide excellent charged particle identification capabilities over a wide momentum range of $2-100$ GeV/c.

The results presented here are based on $6.8~\mu\mathrm{b}^{-1}$ from the LHC 2009 pilot run at $\sqrt{s}$ = 0.9 TeV, $0.3~\mathrm{nb}^{-1}$ at $\sqrt{s}$ = 0.9 TeV accumulated in early 2010 and $14~\mathrm{nb}^{-1}$ accumulated by summer 2010 at $\sqrt{s}$ = 7 TeV. These data-sets were recorded using minimum bias triggers, which require a minimum energy deposit in the LHCb calorimeters or at least one reconstructed track in the event. Due to the finite beam size and crossing angle, the two halves of the VELO were partially retracted during the $\sqrt{s}$ = 0.9 TeV runs.

Soft QCD measurements are important input for modelling of the underlying event and tuning of event generators. Current models have been tuned to describe SPS and Tevatron data, with transverse momentum coverage of $p_T>0.5$ GeV and central rapidity. LHCb has the unique capability to explore uncovered phase space regions. Good understanding of soft QCD processes is also required for extracting many important measurements at the LHC.\\
The measurement presented in this paper are the production cross section of $K_s^0$ at $\sqrt{s}$ = 0.9 TeV, the prompt $\bar{\Lambda}/\Lambda$, $\bar{\Lambda}/K_s^0$ and $\bar{p}/p$ production ratios, the inclusive $\phi$ production cross section at $\sqrt{s}$ = 0.9 TeV and $\sqrt{s}$ = 7 TeV and the measurement of prompt charged particle multiplicity.

\section{Measurement of strangeness production}
Strangeness production studies provide sensitive tests of soft hadronic interactions, as the
mass of the strange quark is of the order of the QCD scale $\Lambda_{QCD}$. Strange-hadron production is suppressed, as a consequence, but still occurs in the non-perturbative regime. Different fragmentation models, tuned on Tevatron data, agree on the total amount of strangeness produced but disagree on its distribution over phase space and on the ratios of the strange hadrons produced. 

LHCb is particularly interesting for QCD studies since one can measure the production of strange quarks in a range where models are expected to diverge more than for central rapidities. Moreover, $V^0$ decays ($K_s^0$, $\Lambda$, $\bar{\Lambda}$) have a very clean experimental signature which allows to identify them unambiguously using only kinematic information. As for the $\phi$ meson, it offers a unique way to study primary strangeness production, and via its decay into charged kaon pairs is vital for the calibration of the PID system of LHCb.

\subsection{$K_s^0$ production cross section}

The hadronic production of $K_s^0$ mesons has been studied by several experiments at a range of different
centre-of-mass energies, both in $pp$ and $p\bar{p}$ collisions (see for example \cite{cdf_k0,star_k0}). The most recent measurements of $K_s^0$ production at the Tevatron have shown deviations with respect to the expectations of hadronization models \cite{cdf_k0}.

This is an ideal early measurement since $K_s^0$ are relatively abundant in minimum bias data and $K_s^0\rightarrow\pi^+\pi^-$ candidates from the 2009 pilot run can be selected using only the tracking system. A high-purity sample could then be obtained without using particle identification information.

\begin{figure}[ht]
\centering
\includegraphics[width=140mm]{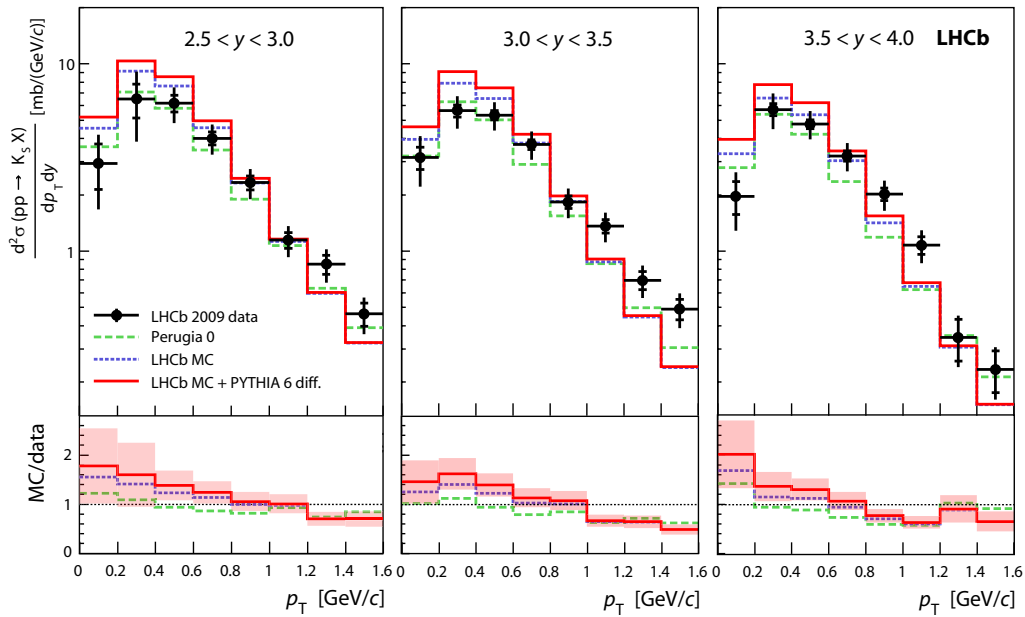}
\caption{Double-differential prompt $K_s^0$ production cross-section in $p\bar{p}$ collisions at $\sqrt{s}$ = 0.9 TeV as a function of transverse momentum $p_T$ and rapidity \textit{y}. The points represent LHCb data. Measurements are compared with the Pythia tune Perugia0 \cite{mctunes} and the LHCb MC with and without diffractive events. The lower plots show the MC/data ratios, with the shaded band representing the uncertainty for one of these ratios.} \label{k0_1}
\end{figure}

In Figure \ref{k0_1}, the LHCb results is presented in bins of \textit{y} and $p_T$. The different Monte Carlo (MC) predictions agree reasonably well with the data, although they tend to underestimate (overestimate) the measured production in the highest (lowest) $p_T$ bins.

Due to the long $K_s^0$ lifetime and partially open VELO position, only a small fraction of the $K_s^0$ daughter tracks traversing the spectrometer leave a signal in the VELO. Therefore, a twofold path was followed for the $K_s^0$ reconstruction and selection, on one hand using tracks reconstructed only with hits in the tracking stations, on the other hand forming $K_s^0$ candidates with tracks leaving hits also in the VELO. The results are then combined. All the efficiencies are estimated using MC simulations.

To determine the absolute luminosity, for this study, a novel method was designed \cite{lhcb_k0, plamen}, employing the high resolution of the LHCb vertex detector to directly measure width and position of the beams using beam-gas and beam-beam interactions. Information about the bunch currents was provided by the LHC machine. In this analysis the main systematic contributions were given by the bunch intensities, 12\%, and by the tracking efficiency, 10\% in the most challenging bins.

\begin{figure}[ht]
\centering
\includegraphics[width=120mm]{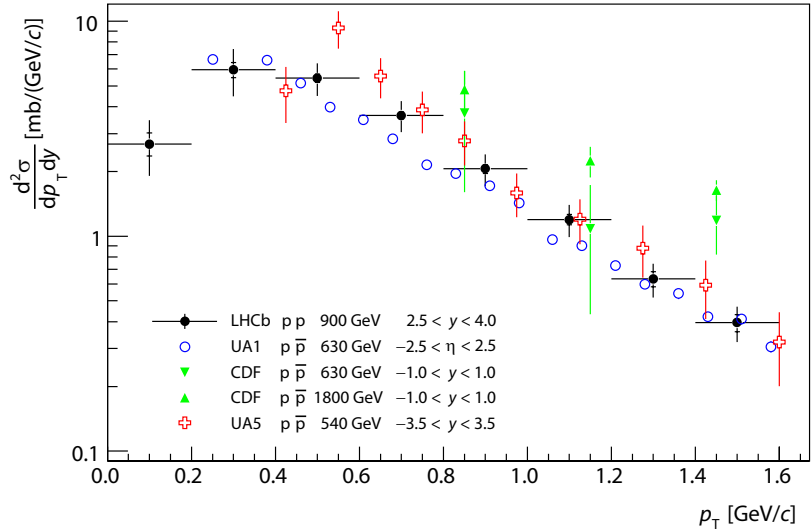}
\caption{Absolute measurements of the $K_s^0$ production cross-section as a function of $p_T$, performed by the UA1, UA5, CDF and LHCb experiments, at different high-energy hadron colliders and in different \textit{y} or $\eta$ ranges.} \label{k0_2}
\end{figure}

The LHCb measurement is the first one at 0.9 TeV. As one can see from Fig. \ref{k0_2}, compared to earlier measurements the range in rapidity, \textit{y}, and transverse momentum, $p_T$ , were extended towards larger and smaller values, respectively. In general the agreement with previous measurements is reasonable, given the range of center-of-mass energies and that the results are averaged over different ranges in rapidity or transverse momentum. A detailed description of the analysis procedure and results are reported in \cite{lhcb_k0}.

\subsection{Inclusive $\phi$ cross section}

A measurement of the inclusive differential $\phi$ cross-section in \textit{pp} collisions at $\sqrt{s}$ = 7 TeV is presented in this study \cite{lhcb_phi}.  $\phi$ mesons are reconstructed using the $K^+K^-$ decay
mode and thus the selection relies strongly on LHCb's RICH detectors for particle identification purposes. 

Two data samples are used, one in which at least one kaon is required to pass a RICH based selection, the tag sample, and one in which both kaons have to pass this selection, the probe sample. To minimize the dependence on MC simulations the PID efficiency in $p_T$ , \textit{y} bins is estimated with this tag and probe technique. The LHCb luminosity used in this cross-section estimation is based on a continuous analysis of hits in the Scintillator Pad Detector \cite{lhcb_descr}, which has been normalized to the absolute luminosity scale. The absolute scale was determined with the beam-gas method mentioned before. The probability for multiple \textit{pp} collisions per bunch-crossing was negligibly low in the data taking period considered.

\begin{figure}[ht]
\centering
\includegraphics[width=84mm]{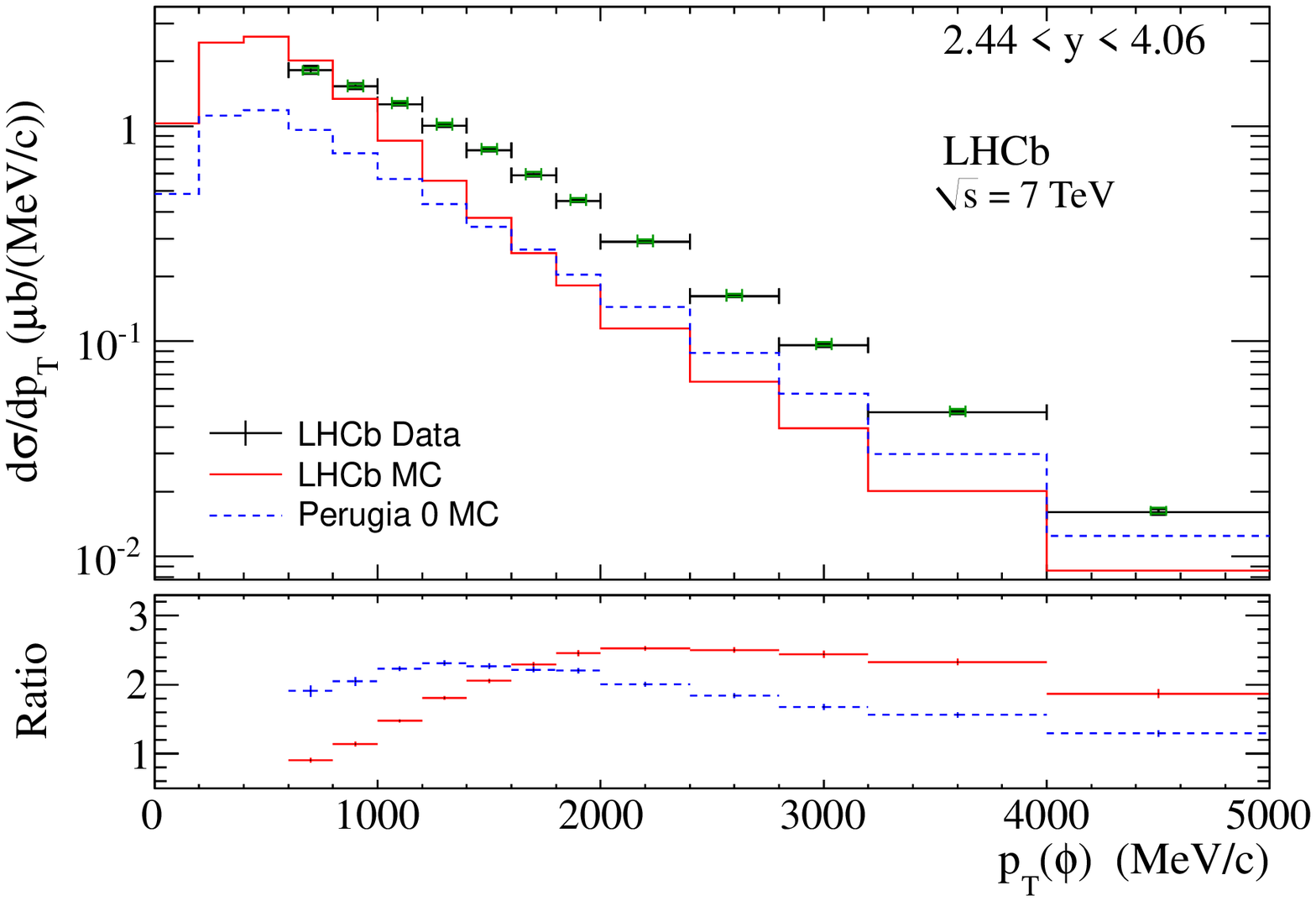}
\includegraphics[width=84mm]{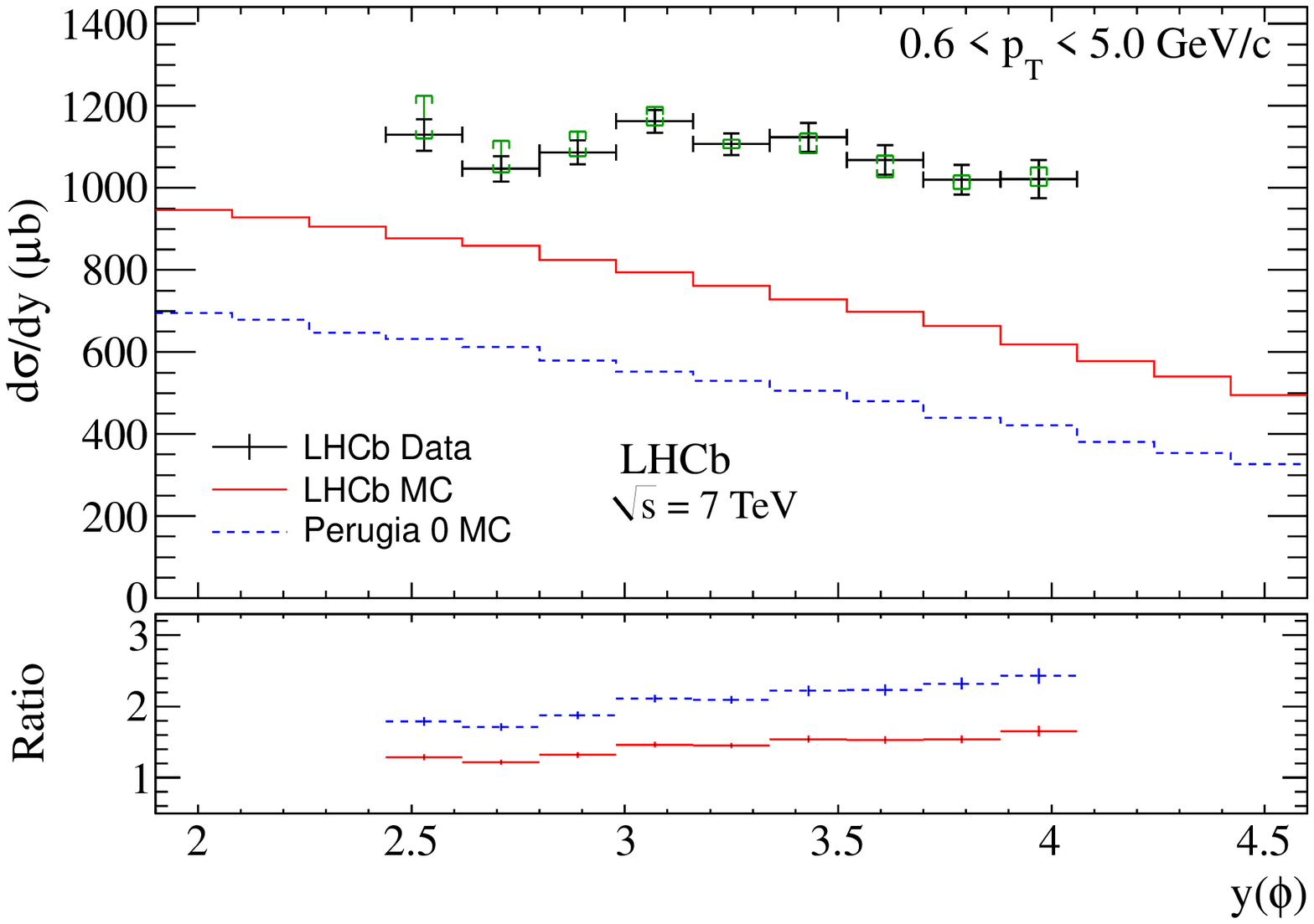}
\caption{Inclusive differential $\phi$ production cross-section as a function of $p_T$ (left)
and \textit{y} (right), measured with data (black), and compared to the LHCb MC (solid, red) and Perugia 0 (dashed, blue). The inner error bars represent the statistical uncertainty, the outer show the quadratic sum of statistical and systematic errors. Error bars shown in the ratio plots are statistical only.} \label{phi_1}
\end{figure}

In Figure \ref{phi_1} are shown projections on the \textit{y} and $p_T$ axis within the same kinematic region. Also plotted are the results from the Perugia 0 \cite{mctunes} tuning of Pythia 6.4 and the standard LHCb MC \cite{lhcbtune} based also on the same version of Pythia. In general, both Monte Carlo samples underestimate the measured $\phi$ meson production in the phase space region of the measurement. There is a tendency to better agreement for lower transverse momenta.
The cross-section for inclusive $\phi$ production in the kinematic range $0.6 < p_T < 5.0\ \mathrm{GeV/}c$ and $2.44 < y < 4.06$ is found to be:
\begin{equation}
\sigma(pp\rightarrow\phi X) = (1758\pm19(stat)~ ^{+43}_{-14} (syst) \pm182(scale))\ \mu \mathrm{b}\ . \nonumber
\end{equation}

\section{Baryon number transport and baryon suppression}

Other probes of strangeness production in high energy hadron collisions are cross-section ratios, where luminosity and many systematic uncertainties cancel. In baryon-antibaryon ratios like $\bar{\Lambda}/\Lambda$ or $\bar{p}/p$, the baryon contains valence quarks in common with the proton, while all three antiquarks of the antibaryon have to be produced in the collision. The ratio of the production cross sections thus measures the baryon-number transport from the beam particles to the final state. Several models exist to describe this transport, but the mechanism is not well understood. Also the strange $\bar{\Lambda}/K^0_s$ ratio is a good test for different fragmentation models, which do not agree between them in this phase space region \cite{muresan} and where there were no experimental data available before.

For the $V^0$ studies, high-purity prompt $K^0_s$ , $\Lambda$ and $\bar{\Lambda}$ samples were selected based on a Fisher discriminant constructed using the logarithms of the impact parameters of the $V^0$ particles and of their daughters \cite{lhcb_v0}. The selection requirement of a primary vertex ensured that mainly the candidates coming from non-diffractive events are kept. For the two available energies the analyzed data correspond to 0.31 $\mathrm{nb}^{-1}$ at $\sqrt{s}$ = 0.9 TeV and 1.8 $\mathrm{nb}^{-1}$ at $\sqrt{s}$ = 7 TeV.

\begin{figure}[!ht]
  \centering
  \subfigure[]{
    \includegraphics[width=0.48\textwidth]{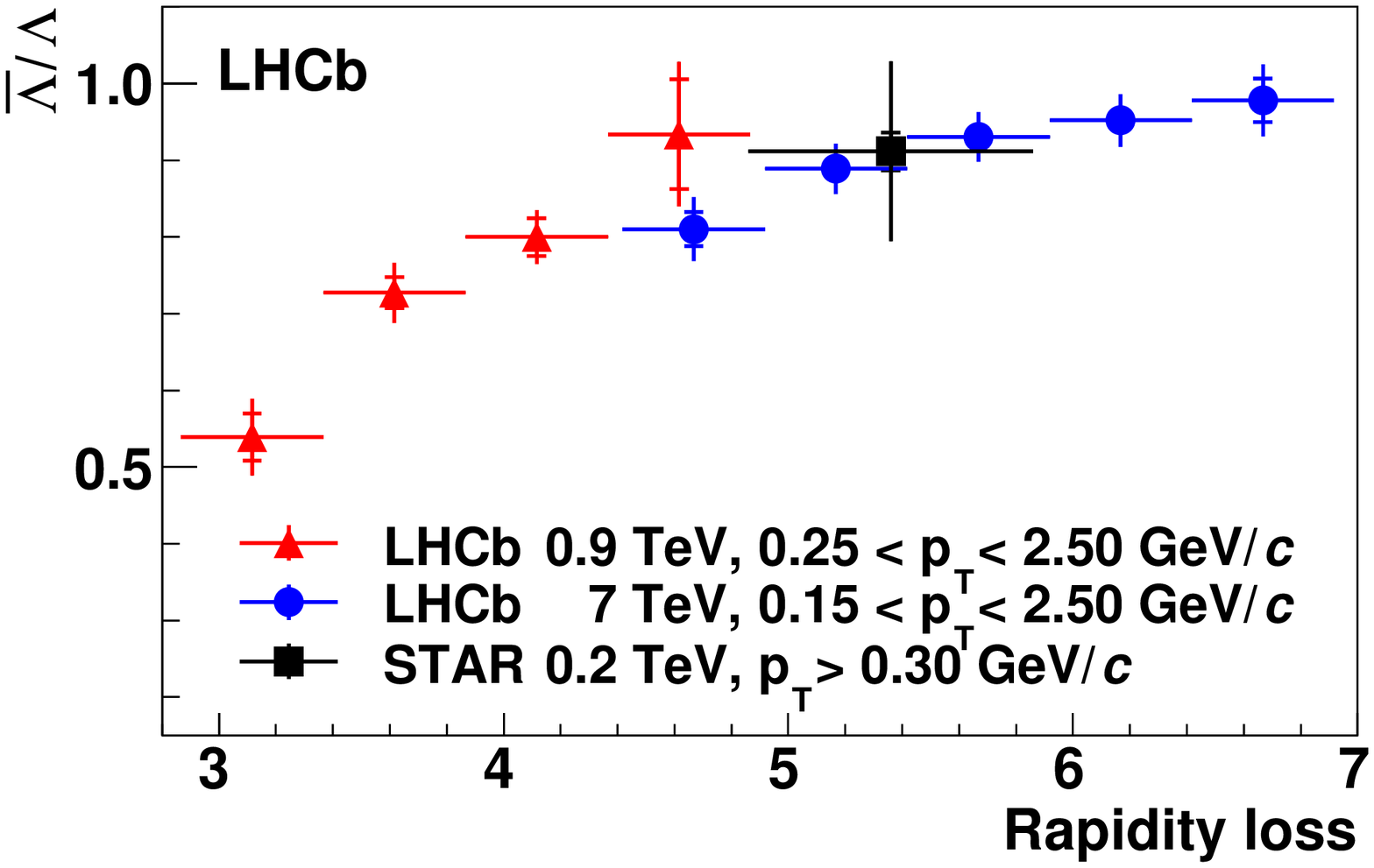}
    \label{fig:plotDeltaY-al}
  }
  \subfigure[]{
    \includegraphics[width=0.48\textwidth]{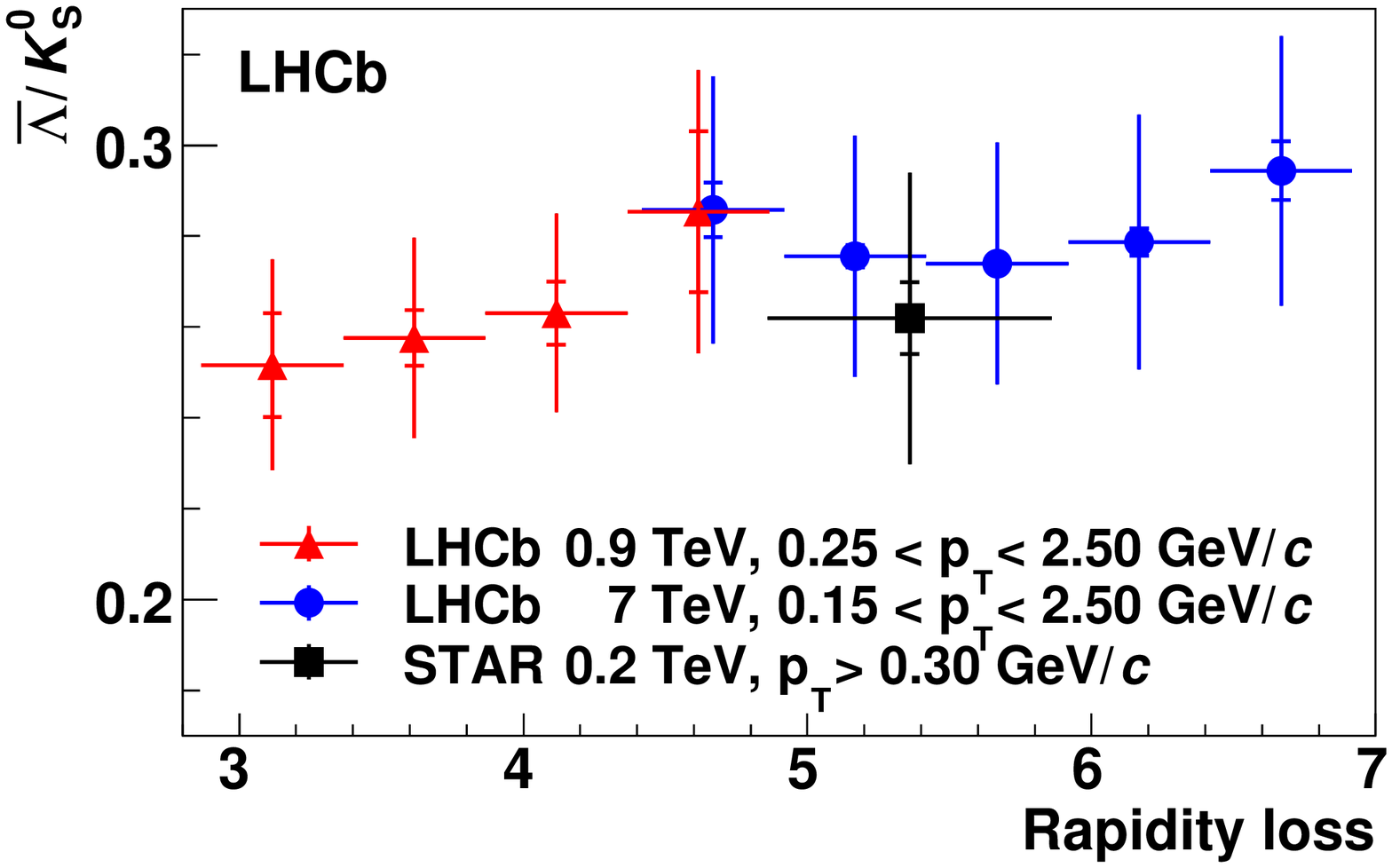}
    \label{fig:plotDeltaY-ak}
  }
  \caption{The ratios (a) $\bar{\Lambda}/\Lambda$ and (b) $\bar{\Lambda}/K^0_s$ from LHCb are compared at both $\sqrt{s}$ = 0.9\,TeV (triangles) and 7\,TeV (circles) with the published results from STAR\,
   \cite{star_k0} (squares) as a function of rapidity loss, $\Delta y = y_\mathrm{beam} - y$.  Vertical lines show the combined statistical and systematic uncertainties and the short horizontal bars (where visible) show the statistical component.}
  \label{v0_1}
\end{figure}

To compare results at both collision energies, and to probe scaling violation, both production ratios are shown in Fig. \ref{v0_1} as a function of rapidity loss, $\Delta y = y_{\mathrm{beam}} - y$\, where $y_{\mathrm{beam}}$ is the rapidity of the protons in the anti-clockwise LHC beam, which travels along the positive \textit{z} direction through the detector. Excellent agreement is observed between results at both $\sqrt{s}$ = 0.9 and 7 TeV as well as with results from STAR at $\sqrt{s}$ = 0.2 TeV. The measured ratios are also consistent with results published by ALICE and CMS.

\begin{figure}[!ht]
  \centering
  \subfigure[]{
    \includegraphics[width=0.48\textwidth]{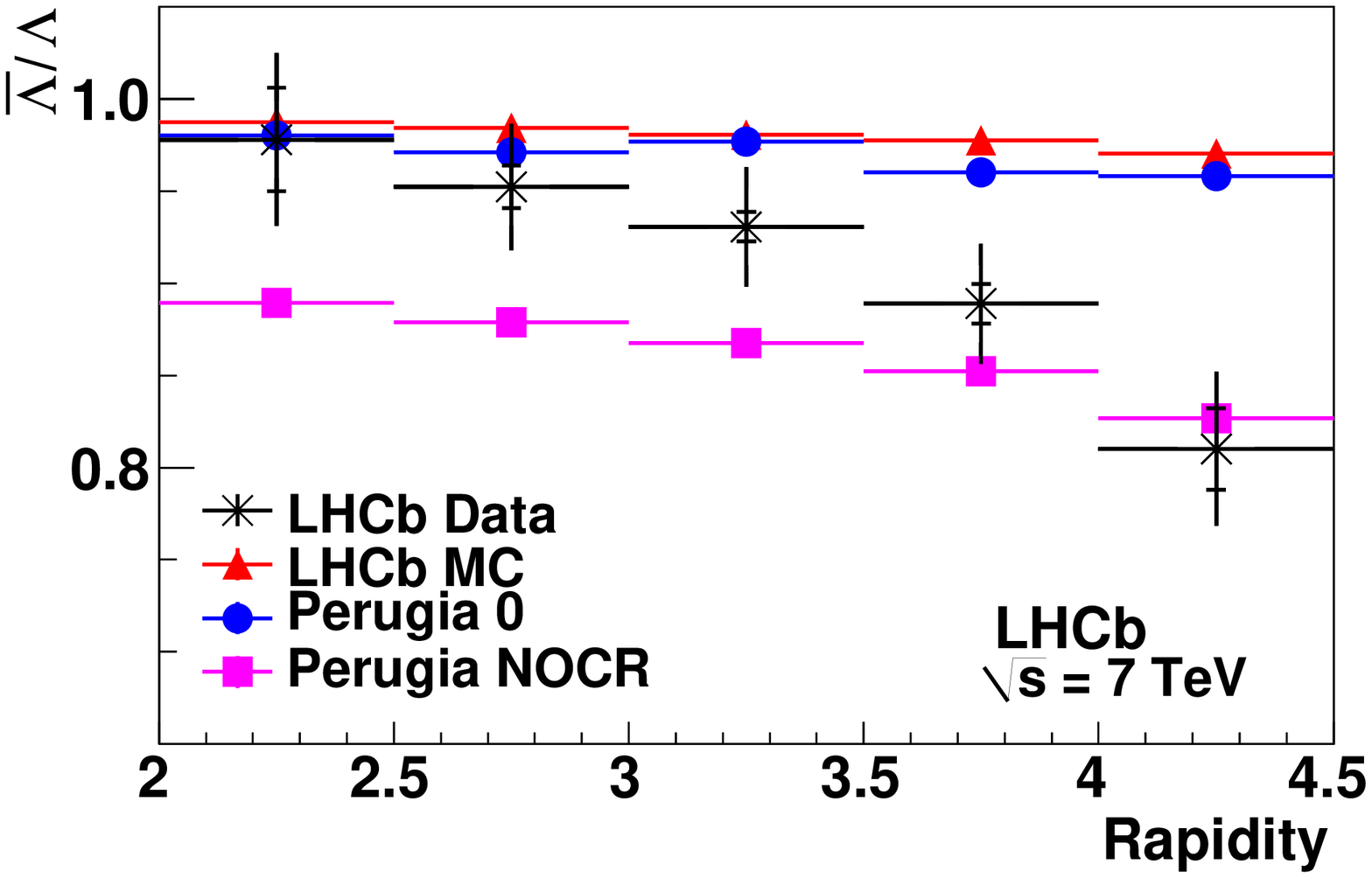}
    \label{fig:plotDataTheory-7-al-y}
  }
  \subfigure[]{
    \includegraphics[width=0.48\textwidth]{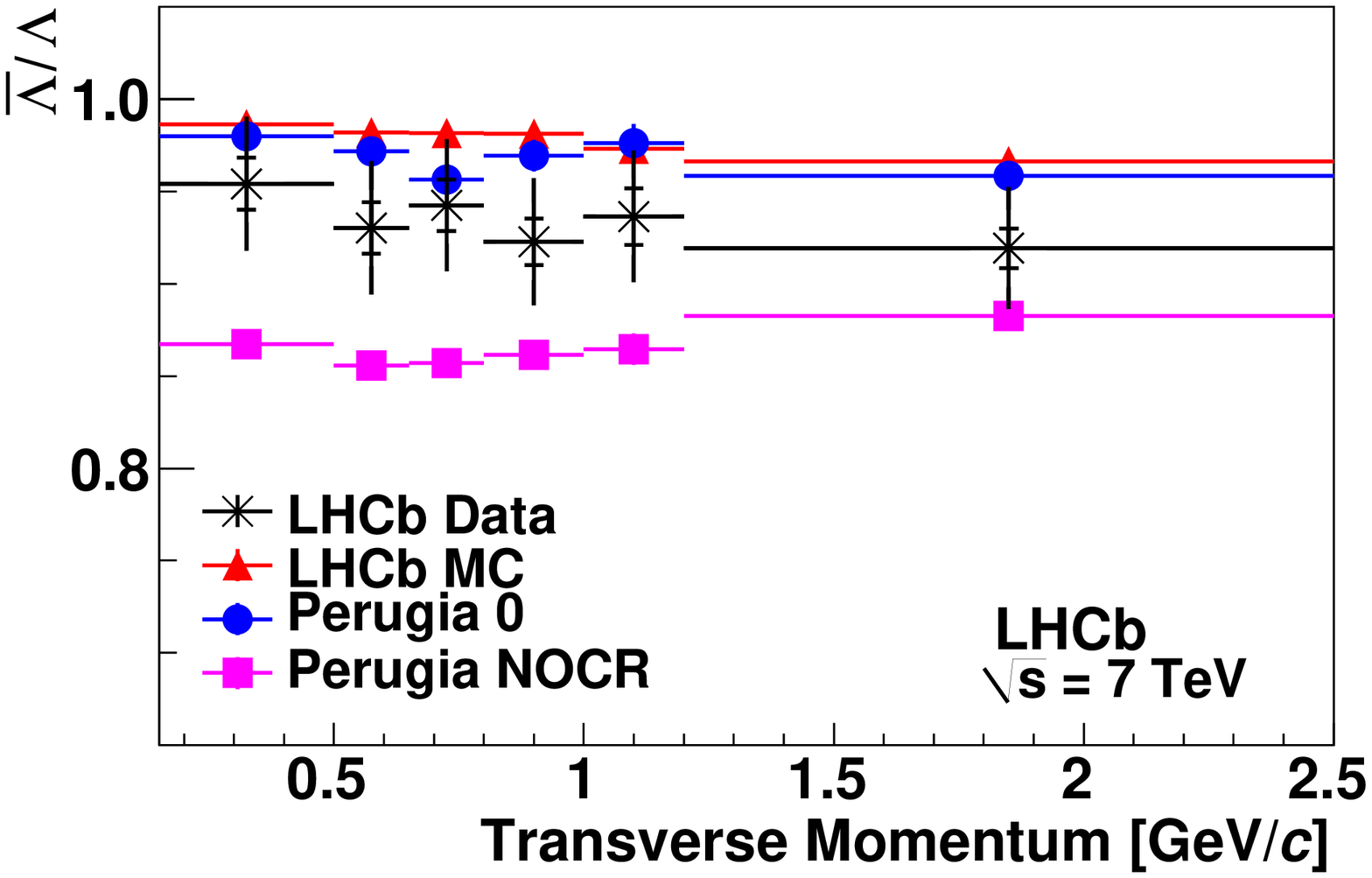}
    \label{fig:plotDataTheory-7-al-pt}
  }
  \subfigure[]{
    \includegraphics[width=0.48\textwidth]{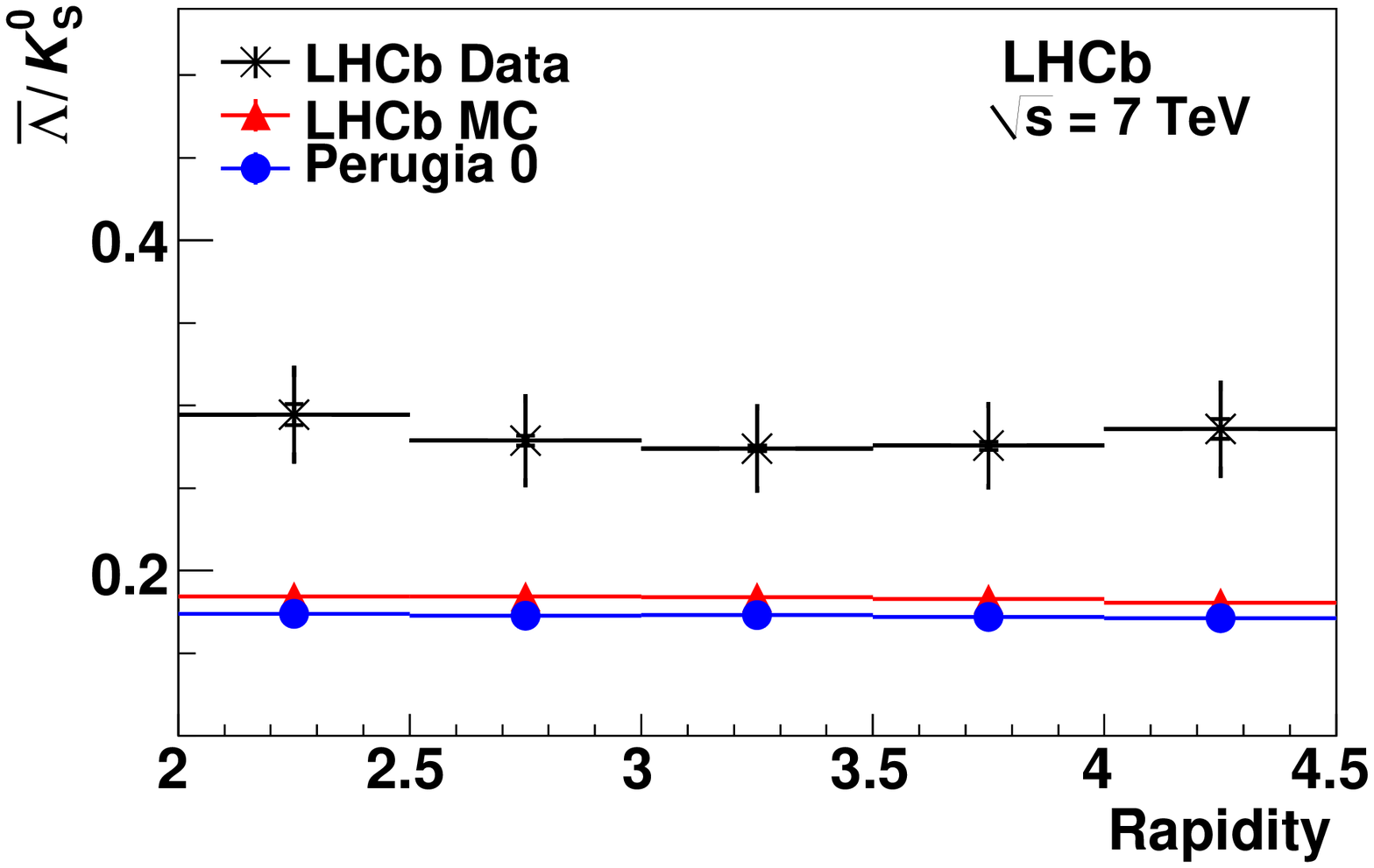}
    \label{fig:plotDataTheory-7-ak-y}
  }
  \subfigure[]{
    \includegraphics[width=0.48\textwidth]{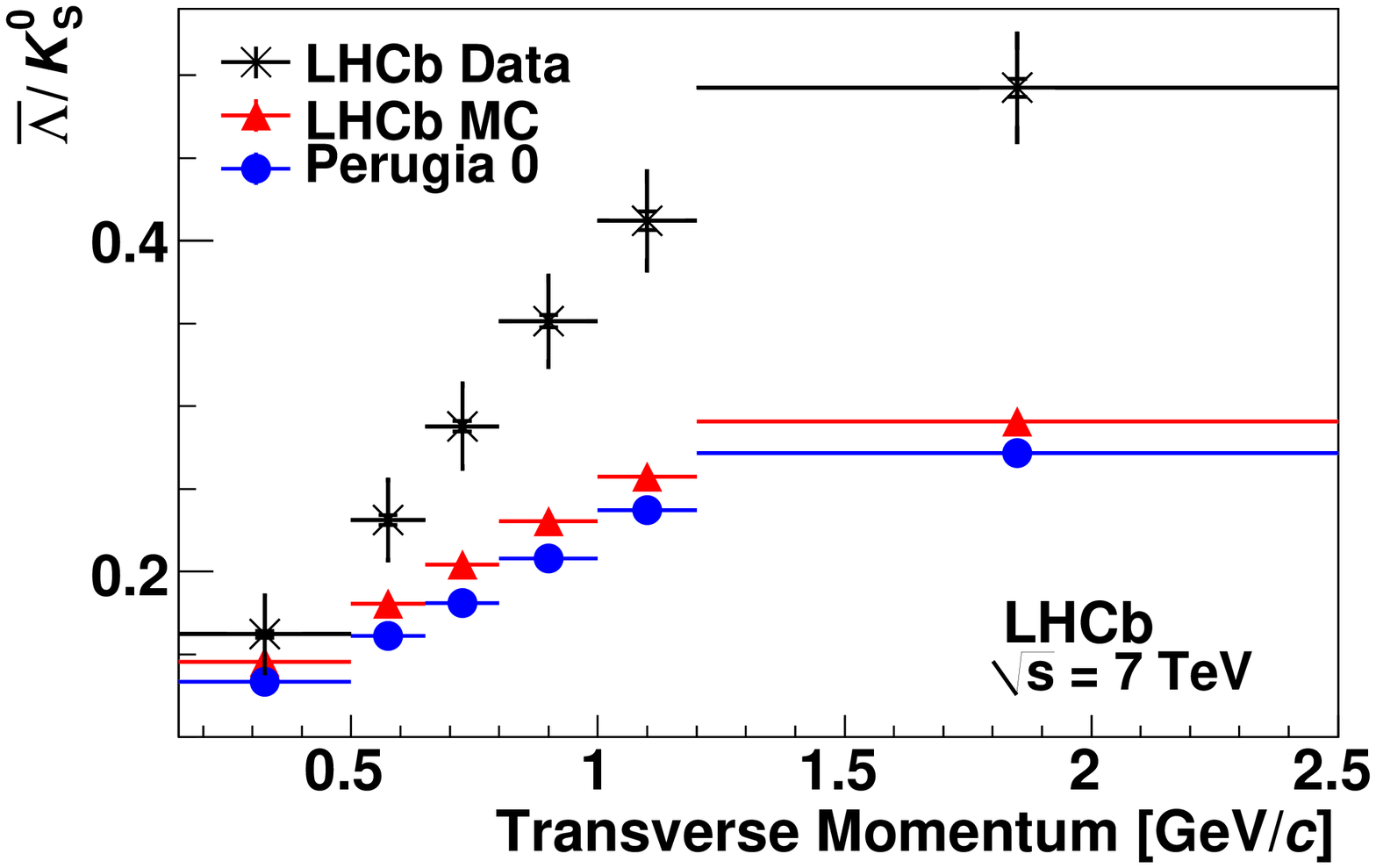}
    \label{fig:plotDataTheory-7-ak-pt}
  }
  \caption{\small The ratios $\bar{\Lambda}/\Lambda$ and $\bar{\Lambda}/K^0_s$ at $\sqrt{s}$ = 7\,TeV compared with the predictions of the LHCb\,MC, Perugia\,0 and Perugia\,NOCR as a function of \subref{fig:plotDataTheory-7-al-y} \& \subref{fig:plotDataTheory-7-ak-y} rapidity and \subref{fig:plotDataTheory-7-al-pt} \& \subref{fig:plotDataTheory-7-ak-pt} transverse momentum.  Vertical lines show the combined statistical and systematic uncertainties and the short horizontal bars (where visible) show the statistical component.}
  \label{v0_2}
\end{figure}

Both measured ratios are compared to the predictions of the Pythia 6 generator tunes: LHCb MC, Perugia 0 and Perugia NOCR \cite{mctunes}, as functions of $p_T$ and \textit{y} at $\sqrt{s}$ = 0.9 and 7 TeV (Fig. \ref{v0_2} shows the latter). According to Monte Carlo studies, the requirement for a reconstructed primary vertex results in only a small contribution from diffractive events to the selected $V^0$ sample, therefore non-diffractive simulated events are used for these comparisons. The predictions of LHCb MC and Perugia 0 are similar
throughout. The ratio $\bar{\Lambda}/\Lambda$ is close to Perugia 0 at low \textit{y} but becomes smaller with higher rapidity, approaching Perugia NOCR. In collisions at $\sqrt{s}$ = 7 TeV, this ratio is consistent with Perugia 0 across the measured $p_T$ range but is closer to Perugia NOCR at $\sqrt{s}$ = 0.9 TeV. The production ratio $\bar{\Lambda}/K^0_s$ is larger in data than predicted by Perugia 0 at both collision energies and in all measurement bins, with the most significant differences observed at high $p_T$\,.

A slight increase in the ratio $\bar{\Lambda}/K^0_s$ is observed when going from $\sqrt{s}$ = 0.9 TeV to 7 TeV. This is plausible, since particle masses and kinematic factors in general should become less important at higher energies. Since both the $\bar\Lambda$ and the $K^0_s$ contain a single strange valence quark, the discrepancy cannot be explained by a mismatch in strangeness suppression between data and Monte Carlo. Instead it probes directly the understanding of baryon production in the fragmentation process.

\begin{figure}[ht]
\centering
\includegraphics[width=170mm]{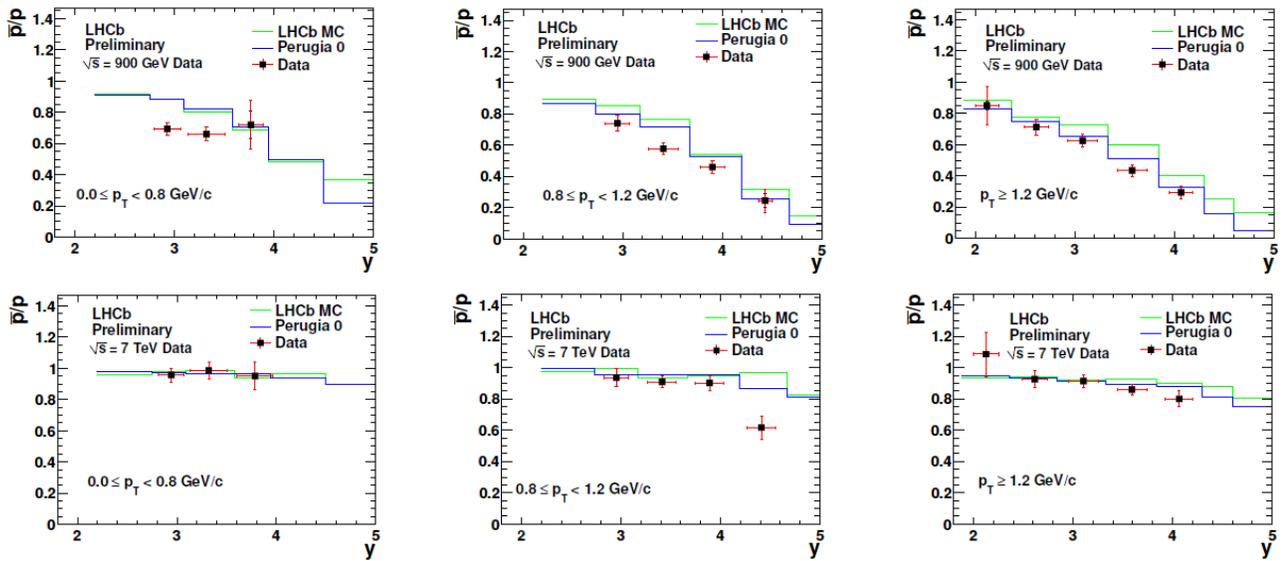}
\caption{Distribution of the ratio $\bar{p}/p$ against \textit{y}: up at $\sqrt{s}$ = 0.9\,TeV, down at $\sqrt{s}$ = 7\,TeV.
} \label{pp_1}
\end{figure}

\begin{figure}[ht]
\centering
\includegraphics[width=120mm]{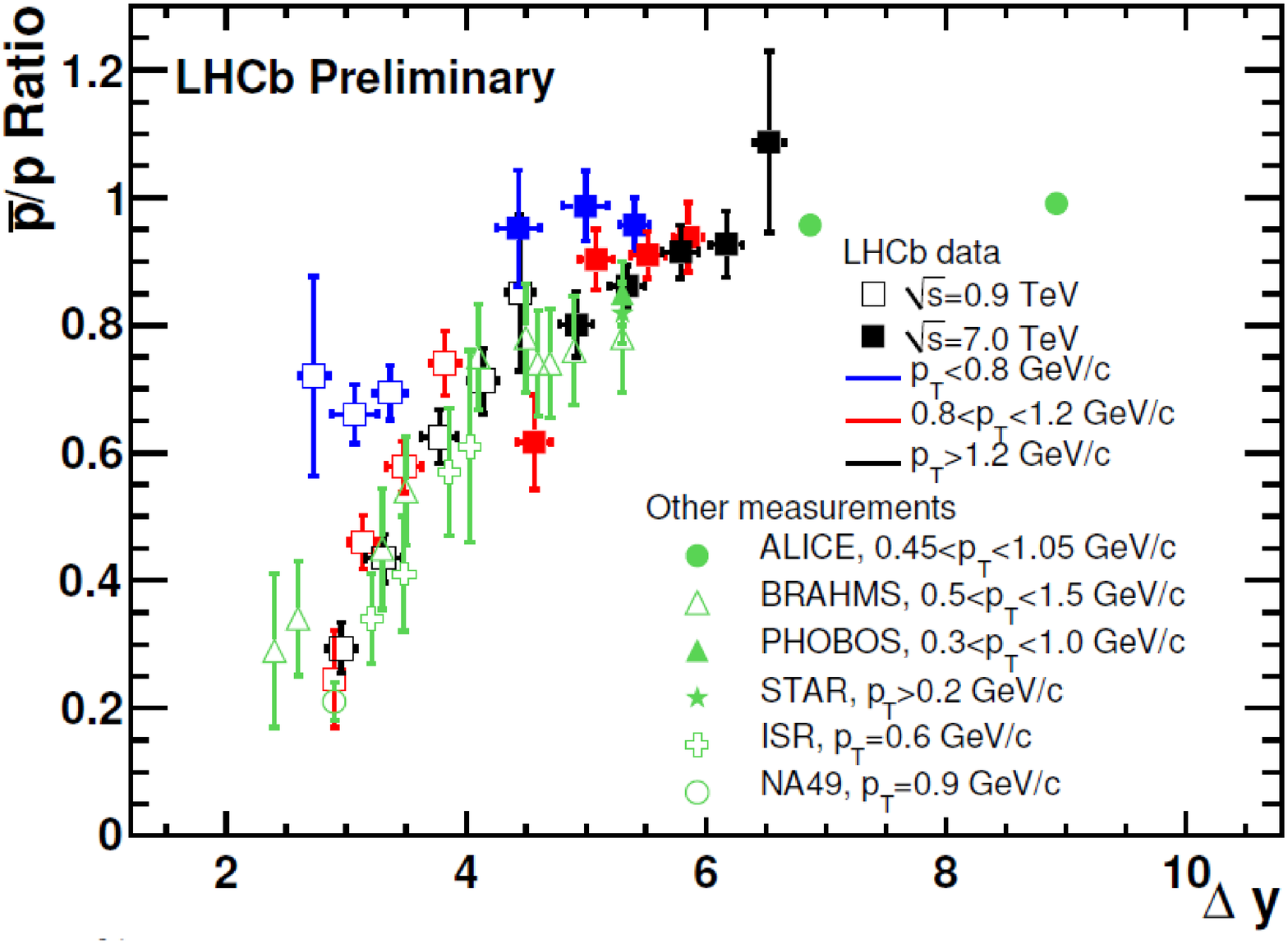}
\caption{LHCb measurements of the $\bar{p}/p$ ratio plotted versus $\Delta y$.} \label{pp_2}
\end{figure}

For the measurement of $\bar{p}/p$ ratio a data sample similar to the previous study was used \cite{lhcb_pp}. RICH information allows to isolate almost pure samples of pions, kaons and protons using delta log-likelihood cuts on particle type hypotheses. The hadron particle identification was calibrated using secondaries from
the decays $K^0_s\rightarrow\pi^+\pi^-$ and $\Lambda\rightarrow p\pi^-$, with pions and protons identified
using kinematic cuts only. Charged kaons were selected from the decay $\phi\rightarrow K^+K^-$ using a tag and probe approach where only one of the kaons was required to be identified by RICH. Similar to the baryon transport pattern for $\bar{\Lambda}/\Lambda$, the $\bar{p}/p$ ratio shows a big deviation in ratio from unity at low energy (except for the highest $p_T$ bin), Fig. \ref{pp_1}. These results are compared against the predictions of Pythia as implemented in the LHCb simulation and those of the so-called Perugia 0 tune of Pythia \cite{mctunes}. In this study LHCb Pythia includes the simulation of diffractive events, whereas Perugia 0 does not. For $\sqrt{s}$ = 0.9 TeV at high $p_T$ there is good agreement between the data and the Perugia 0 predictions, but less so for LHCb Pythia; in the low and middle $p_T$ bins at this energy the data values are in general below those of both generators. At $\sqrt{s}$ = 7 TeV the measurements are compatible with the expectation of both LHCb Pythia and Perugia 0. The results at both energies can also be considered in terms of the rapidity loss $\Delta y = y_\mathrm{beam} - y$, the measurements are shown in this form in Fig. \ref{pp_2}. These span an interval in $\Delta y$ of almost 4 units, reaching up to a highest value of close to $\Delta y=7$. In a given $p_T$ bin the measurements appear consistent with a monotonic distribution. There is an indication that the results have some $p_T$ dependence. Figure \ref{pp_2} also shows measurements of the same quantity performed by other experiments. Reasonable consistency is observed.

\section{Measurement of Charged Particle Multiplicities}

The charged particle production in proton-proton collisions at a centre-of-mass energy of
 $\sqrt{s}$ = 7 TeV in different bins of pseudo-rapidity $\eta$ have been studied with the LHCb detector and compared with model predictions. The charged particles are reconstructed in the VELO detector close to the interaction region, essentially outside of the magnetic field, which provides high reconstruction efficiency in the $\eta$ ranges $-2.5 < \eta < -2.0$ and $2.0 < \eta < 4.5$. In the absence of almost any magnetic field in the VELO region, no further acceptance corrections as function of momentum are needed. Since the VELO detector is close to the interaction region, the amount of material before the particle detection is minimal, minimizing the corrections for particle interactions with detector material. The data were taken with a minimum bias trigger, only requiring a minimum of one reconstructed track in the VELO detector.
 
 Monte Carlo event simulation is used to correct for acceptance and resolution effects. The detector simulation is based on the Geant 4 \cite{geant} program and incorporates our best knowledge of the LHCb spectrometer. The Monte Carlo event samples were passed through reconstruction and selection procedures identical to those for the data. Proton-proton collisions, both elastic and inelastic processes, were generated using the Pythia 6.4 event generators which was tuned to lower energy hadron collider data \cite{lhcbtune}.

The reconstructed multiplicity distributions are corrected on an event by event basis to account for the tracking and selection efficiencies and for the background contributions. These corrected distributions are then used to measure the charged particles multiplicities in each of the $\eta$ ranges though an unfolding procedure. The distributions are corrected for pile-up event. Hard interactions can be enriched by requiring at least one high $p_T > 1 \mathrm{GeV}/c$ track in the forward acceptance of the event. The acceptance is no longer independent of the momentum and requires an additional correction to take into account the efficiency to reconstruct a high $p_T$ track.

The LHCb Monte Carlo is used to estimate the overall tracking and selection efficiency as a function of pseudorapidity and azimuthal angle $\phi$. The main correction come from non-prompt particle contamination (5-10\%), mainly tracks from converted photons.

\begin{figure}[ht]
 \centering
    \subfigure[\ Pythia 6, Pythia 8 and PHOJET]{
    \includegraphics[width=0.47\textwidth]{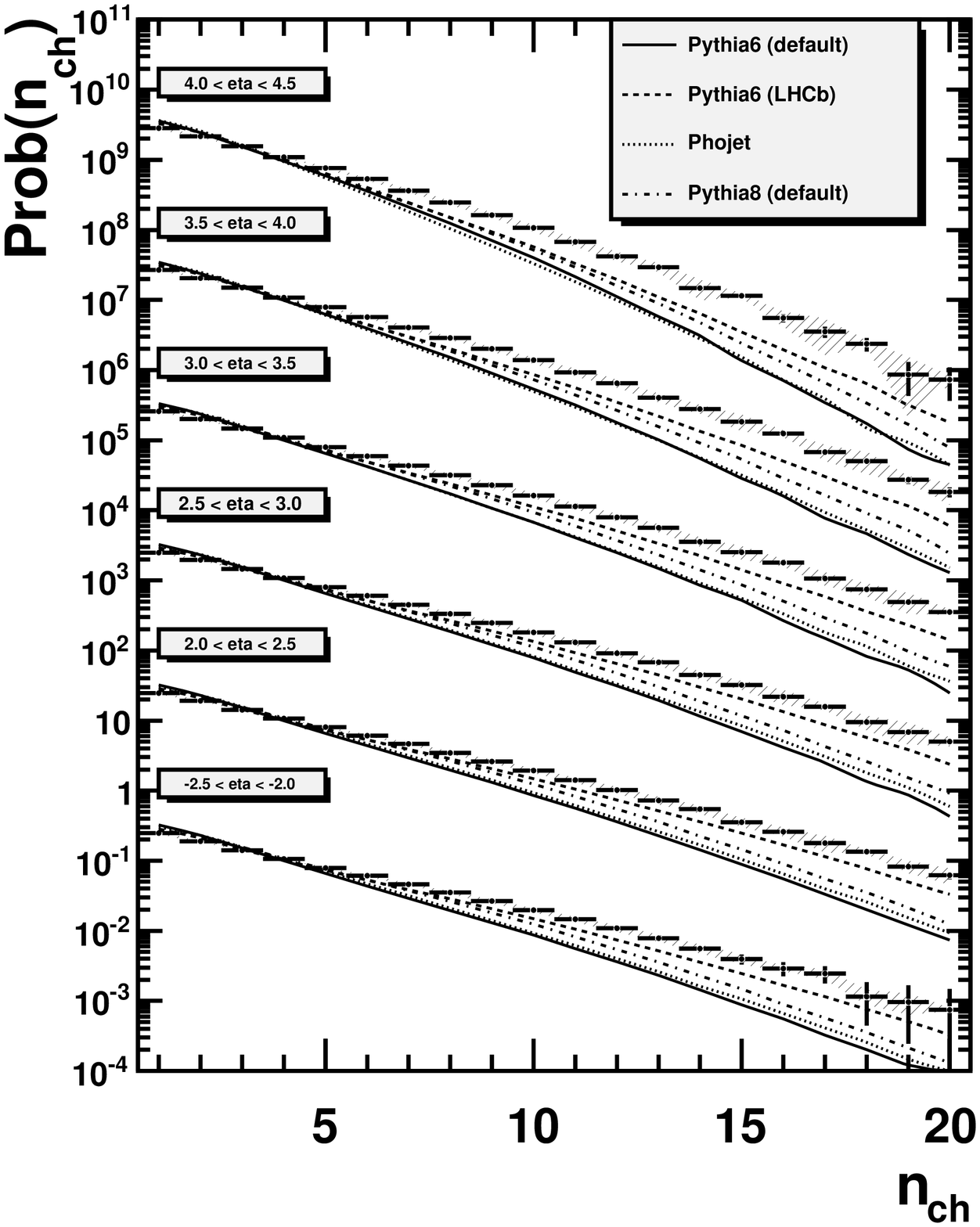}
    \label{part1}
  }
    \subfigure[\ Perugia tunes of Pythia 6 with and without diffraction]{
    \includegraphics[width=0.47\textwidth]{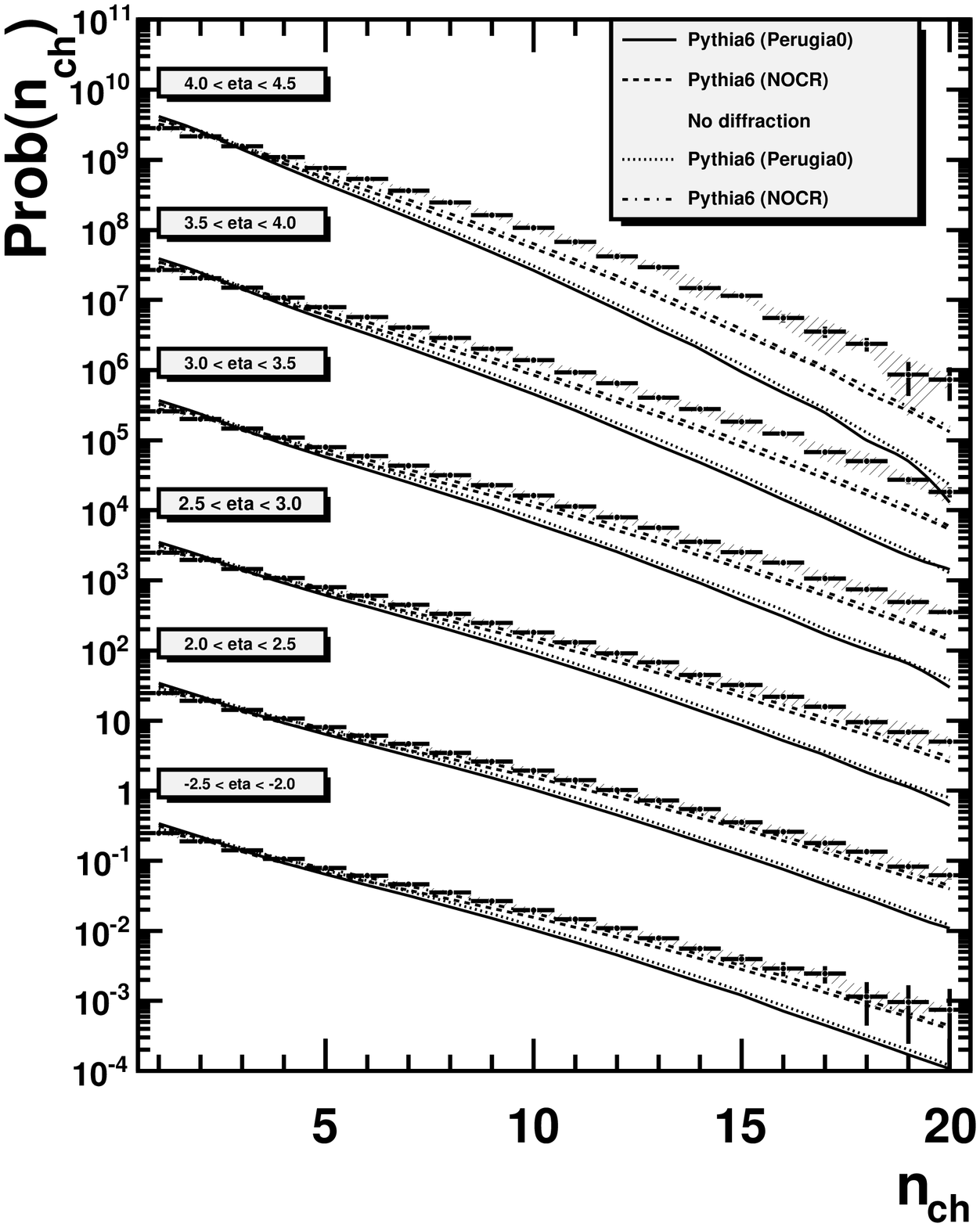}
    \label{part2}
  }
  \caption{Comparison of the LHCb data (points with error bars) with predictions of different  event generators. The different $\eta$-distributions are scaled by factors of $10$ to fit in one plot. The shaded bands represents the systematic uncertainty.}   \label{mult_1}
\end{figure}

Fig. \ref{mult_1} shows the charged particle multiplicity distribution for different ranges
in pseudorapidity $\eta$. In general all generators underestimate the multiplicity distributions, with the LHCb tune giving the best description of the data. The exclusion of the Pythia diffractive processes in the Perugia tunes also improves the description of the data, particularly in the full forward region.

\begin{figure}[ht]
  \centering
     \subfigure[\ Pythia 6, Pythia 8 and PHOJET]{
     \includegraphics[width=0.43\textwidth]{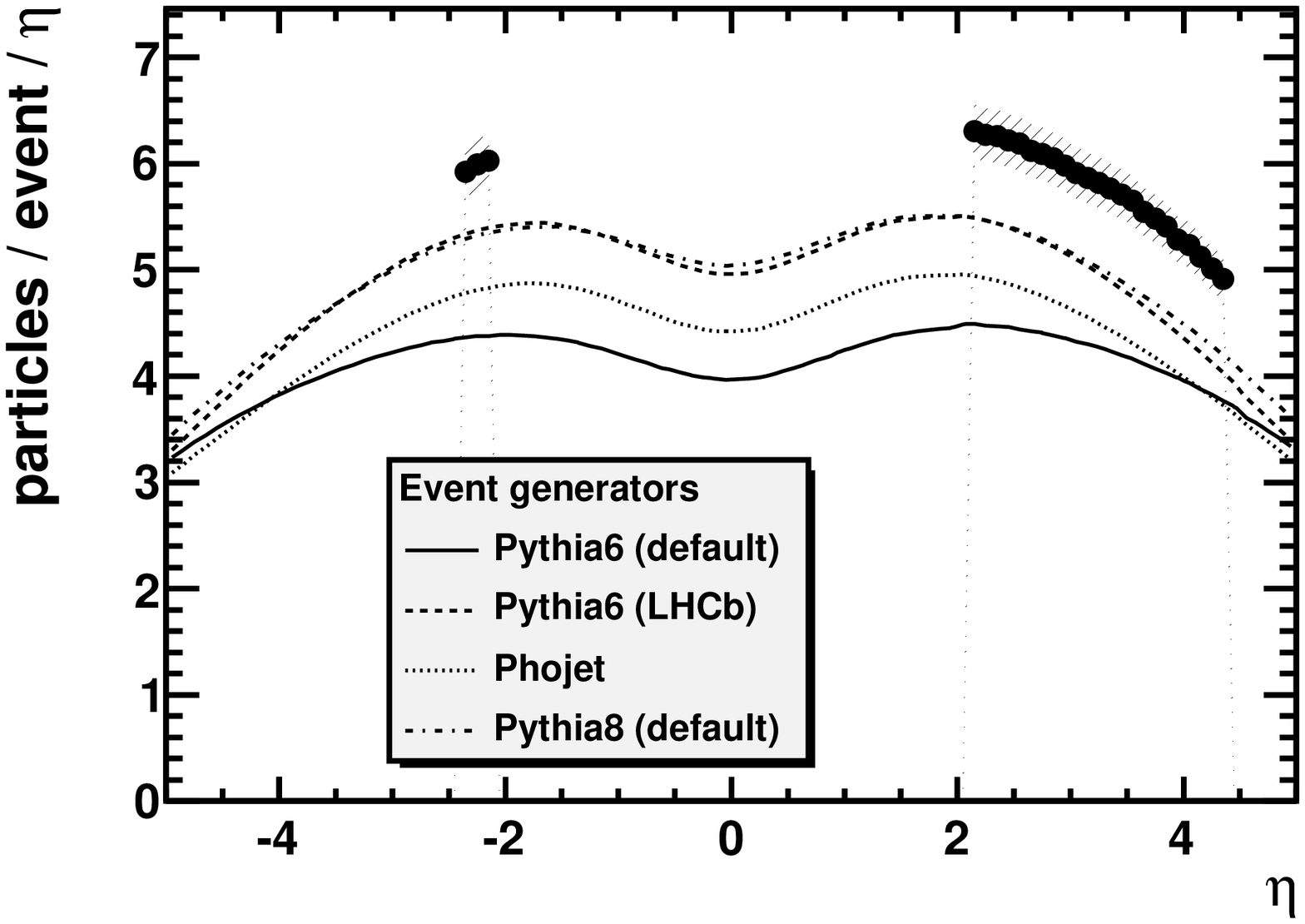}
  }
     \subfigure[\ Perugia tunes of Pythia 6 with and without diffraction]{
     \includegraphics[width=0.43\textwidth]{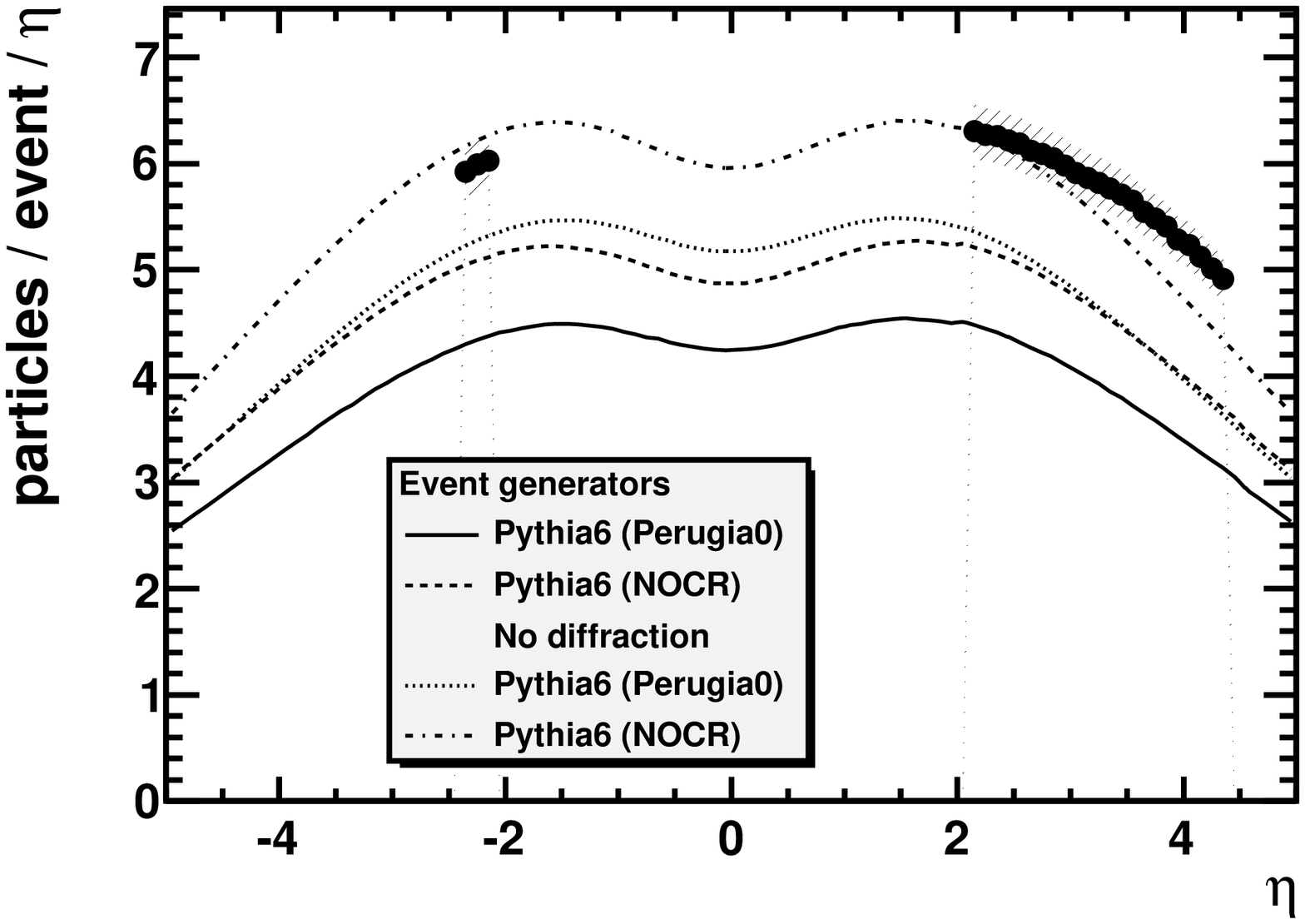}
  }
  \caption{The LHCb data charged particle densities (points) as a function of $\eta$ and comparisons with predictions of event generators.}
  \label{mult_2}
\end{figure}

The charged particle density as a function of pseudorapidity is shown in Fig. \ref{mult_2}. The data has a marked asymmetry between the forward and backward region; this is a consequence of the requirement of at least one track in the full forward η range. All models fail to describe the mean charge particle multiplicity per unit of pseudorapidity.

\section{Conclusions}

The first studies with data from the 2009 and 2010 runs showed that the
LHCb experiment is ready for its core physics program. Early data were
used to perform and to check the calibration of the different subdetectors,
as well as to perform new measurements in a unique range of rapidity and
transverse momentum. A first $K^0_s$ cross-section measurement at 0.9 TeV
was produced at larger rapidities and smaller transverse momenta than
previous measurements. Preliminary results for $\eta$ inclusive cross section
were shown. Preliminary measurements for ratios of $V^0$ and protons suggest
lower baryon suppression and higher baryon transport in data than in the
Monte Carlo models investigated. The LHCb data are consistent with data
from lower energy experiments. In general, the models underestimate the charged particle production, LHCb has already provided great input for their improvement.


\end{document}